\documentclass[a4paper]{ESASPCS13Style}
\usepackage{epsfig}

\begin{document}

\title{The ELODIE Planet Search: Synthetic View of the Survey and its Global Detection Threshold}

\author{D. Naef\inst{1,2} \and M. Mayor\inst{2} \and J.-\,L. Beuzit\inst{3} \and C. Perrier\inst{3} \and D. Queloz\inst{2} 
\and J.-\,P. Sivan\inst{4} \and S. Udry\inst{2}} 

\institute{European Southern Observatory, Alonso de Cordova 3107, Santiago 19, Chile 
      \and Observatoire de Gen\`eve, 51 ch. des Maillettes, CH--1290 Sauverny, Switzerland
      \and Laboratoire d'Astrophysique de Grenoble, UJF, BP 53, F--38041 Grenoble, France
      \and Laboratoire d'Astrophysique de Marseille, Traverse du Siphon, BP 8, F--13376 Marseille, France}

\maketitle 

\begin{abstract}
In this paper, we give a synthetic view of the {\footnotesize ELODIE} Planet Search programme: a short description 
of our instrument and the surveyed sample as well as a brief review of our detections.
Moreover, we have obtained, through numerical simulations, the global survey sensitivity: a detection probability 
map in the $m_{\rm 2}$ versus $P$ diagram. We use this map for correcting our total number of detections for 
observational biases. Finally we derive the fraction of our sample stars hosting at least one giant planet.

\end{abstract}

\section{The ELODIE survey: quick view}
\label{quick view}
The search for extra-solar planets with the {\footnotesize ELODIE} echelle spectrograph (\cite{Baranne96}) mounted 
on the 193--cm telescope at Observatoire de Haute-Provence ({\footnotesize OHP}) started in 1994. The initial sample 
contained 142 stars, out of which \object{51 Peg} (\object{HD 217014}), the star hosting the first detected 
extra-solar planet (\cite{MQ95}). The sample was largely modified in 1997. The to-date survey sample size amounts to 
330 stars. 18 extra-solar planet candidates have been detected with {\footnotesize ELODIE}. 15 of these candidates 
are orbiting a star in our sample. The three other detections (\object{Gl 876 b}, \object{HD 80606 b} and 
\object{HD 178911 Bb}) result from other programmes. 
Here are the main characteristics of our survey:

\begin{table}[!htp]
  \caption{The 18 objects with minimum masses below 18 M$_{\rm Jup}$ detected with {\footnotesize ELODIE}
  and planet candidates confirmed with this instrument}
  \label{table1}
  \begin{center}
    \leavevmode
    \footnotesize
    \begin{tabular}[h]{ll}
      \hline \\[-5pt]
      Planet & Reference\\[+5pt]
      \hline \\[-5pt]
      \object{51 Peg b}        & \cite{MQ95}\\
      \object{14 Her b}        & \cite{Naef04}\\
      \object{Gl 876 b}        & \cite{Delfosse98}\\
      \object{HD 209458 b}     & \cite{Mazeh00}\\
      \object{HD 190228 b}     & \cite{Perrier03}\\
      \object{HD 8574 b}       & \cite{Perrier03}\\
      \object{HD 50554 b}      & \cite{Perrier03}\\
      \object{HD 74156 b}      & \cite{Naef03}\\
      \object{HD 74156 c}      & \cite{Naef03}\\
      \object{HD 80606 b}      & \cite{Naef01}\\
      \object{HD 106252 b}     & \cite{Perrier03}\\
      \object{HD 178911 Bb}    & \cite{Zucker02}\\
      \object{HD 20367 b}      & \cite{Udry02}\\
      \object{HD 23596 b}      & \cite{Perrier03}\\
      \object{HD 33636 b}      & \cite{Perrier03}\\
      \object{HD 37124 c}      & \cite{Udry02}\\
      \object{HD 150706 b}     & \cite{Udry02}\\
      \object{Gl 777 Ab}       & \cite{Naef03}\\[+5pt]
      \hline
      \multicolumn{2}{c}{Confirmations}\\
      \hline \\[-5pt]
      \object{Ups And b}      & \cite{Naef04}\\
      \object{Ups And c}      & \cite{Naef04}\\
      \object{Ups And d}      & \cite{Naef04}\\
      \object{55 Cnc b}       & \cite{Naef04}\\
      \object{55 Cnc d}       & \cite{Naef04}\\
      \object{47 UMa b}       & \cite{Naef04}\\
      \object{70 Vir b}       & \cite{Naef04}\\
      \object{HD 187123 b}    & \cite{Naef04}\\
      \hline \\
      \end{tabular}
  \end{center}
\end{table}
  
\begin{figure}[!t]
  \begin{center}
    \epsfig{file=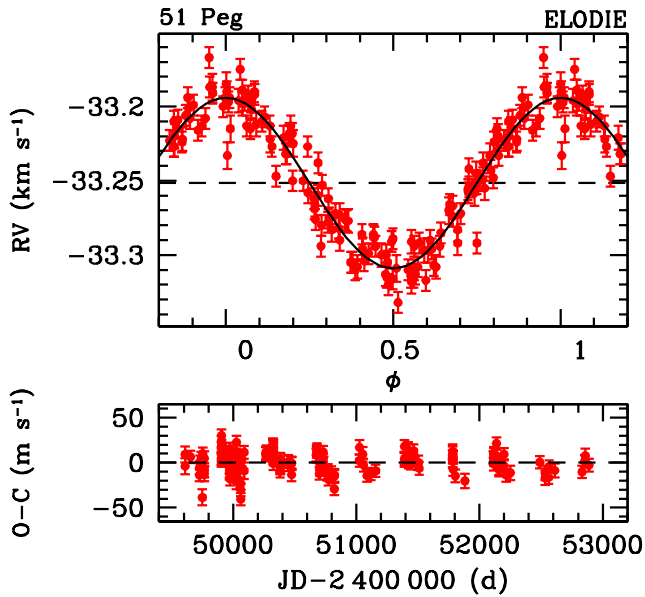, width=0.95\hsize}
  \end{center}
  \caption{3000 days of 51 Peg radial-velocity  follow-up. Top: phase-folded ELODIE velocities and fitted orbital 
  solution. Bottom: residuals to the fitted orbit. Figure from Naef et al. 2004.\label{fig1}}
\end{figure}

\begin{itemize}
  \item {\footnotesize\bf ELODIE:} $\frac{R}{\Delta R}$\,=\,42\,000 echelle spectrograph mounted on the 193--cm 
  telescope at {\footnotesize OHP} ({\footnotesize CNRS}, France). A detailed description of the instrument can 
  be found in \cite{Baranne96}.
  
  \item {\bf Instrumental precision:} $\simeq$\,6.5\,m\,s$^{\rm -1}$ (see \cite{Perrier03}) using the 
  {\sl simultaneous Thorium-Argon} technique.
  
  \item {\bf Sample:} 330 solar-type stars brighter than $m_{\rm V}$=7.65 in the northern hemisphere. The fast 
  rotators ($v\sin i$ $>$ 4\,km\,s$^{\rm -1}$) and the binaries were removed according to {\footnotesize CORAVEL} 
  (\cite{Baranne79}) radial-velocity data (see \cite{Perrier03} for details).
  
  \item {\bf Detections:} 18 planets detected (15 within the above described planet-search sample). Some of these 
  planets are in  multiple systems: \object{HD 37124 c} (\cite{Udry02}); \object{HD 74156 b} and c (\cite{Naef04}). 
  The complete list of the to-date {\footnotesize ELODIE} detections and the corresponding references are presented 
  in table~\ref{table1}. As an example of detection,  Fig.~\ref{fig1} shows the {\footnotesize ELODIE} updated 
  orbital solution for \object{51 Peg} published in \cite{Naef04}. 
  
  \item {\bf Confirmations:} Using {\footnotesize ELODIE}, we have confirmed the orbital solution for planet 
  candidates around \object{Ups And} (\object{HD 9826}, \cite{Butler97}, \cite{Butler99}) \object{55 Cnc} 
  (\object{HD 75732}, \cite{Butler97}, \cite{Marcy02}), \object{47 UMa} 
  (\object{HD 95128}, \cite{Butler96}), \object{70 Vir} (\object{HD 117176}, \cite{Marcy96}) and \object{HD 187123} 
  (\cite{Butler98}). These confirmed planetary companions are also listed in table~\ref{table1}.

\end{itemize}

\section{Global {\footnotesize ELODIE} survey sensitivity}

We have determined, via numerical simulations, the global {\footnotesize ELODIE} survey sensitivity, i.e. the 
probability of detection in the secondary mass versus orbital period diagram. We give some details about these 
simulations in Sect.~\ref{simul}. In Sect.~\ref{jitter}, we describe how we accounted for non-photonic error sources. 
We study the impact of various stellar properties on the sensitivity in Sect.~\ref{bvfeh}. Finally, we present our 
results in Sect.~\ref{results}.

\subsection{Numerical simulations}
\label{simul}

We have computed, through numerical simulations, the detection probabilities for a grid in the $m_{\rm 2}$ versus $P$ 
diagram. The secondary mass and orbital period considered intervals are: 
0.025\,$\leq$\,$m_{\rm 2}$\,$\leq$\,20\,M$_{\rm Jup}$ and 0.8\,$\leq$\,$P$\,$\leq$\,6000\,d. The total number of grid 
points is 3534. We have generated 5000 random orbits for each grid point using the following distributions:

\begin{itemize}
  \item uniform distribution for $T_{\rm 0}$ (instant of periastron passage) and $\omega$ (longitude of the periastron)
  
  \item $e$ (orbital eccentricity): we have used the eccentricity distribution observed for the known extra-solar planet 
  candidates
  
  \item $i$ (inclination of the orbital plane): the probability density we have used for $i$ is proportional to 
  $\sin i\,di$
\end{itemize}

\vspace{0.25cm}
\noindent
Our simulation algorithm accounts for:

\vspace{0.25cm}
\begin{itemize}
  \item the stellar content of our sample (i.e. the $M_{\rm 1}$, $m_{\rm V}$, $B-V$, $[$Fe/H$]$ and $v\sin i$ 
  distributions for our sample)
  
  \item the real timing of the radial-velocity observations
  
  \item weather and seeing conditions at {\footnotesize OHP}
  
  \item the presence of non-photonic error sources such as stellar activity jitter (see Sect~\ref{jitter})
\end{itemize}

\subsection{Non-photonic error sources}
\label{jitter}

In order to account for the presence of non-photonic velocity error sources, we have determined the distribution of the 
observed velocity dispersion corrected for the photon noise for a subsample of target stars. This subsample contains the 
non-variable stars and the micro-variable stars for which the variability origin is unknown. The stars with planetary 
companions, the spectroscopic binaries ({\footnotesize SB1} or {\footnotesize SB2}), the stars with close visual companions 
and the stars with blended spectra have been removed. 
 
\begin{figure}[!t]
  \begin{center}
    \epsfig{file=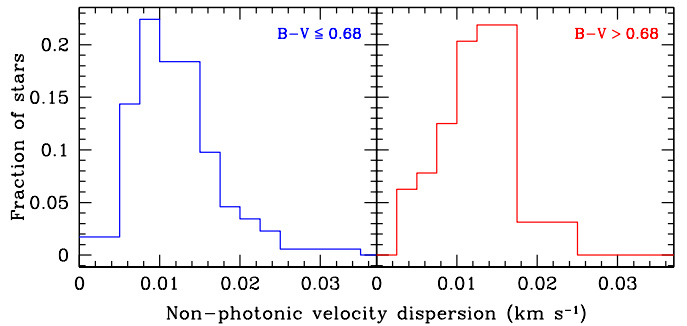, width=\hsize}
  \end{center}
  \caption{Distributions of non-photonic error sources obtained for the {\footnotesize ELODIE} planet search sample. Two 
  colour intervals have been considered: $B-V\leq$0.68 (left) and $B-V>$0.68 (right). We see that the non-photonic noise 
  only weakly depends on star colours.\label{fig2}}
\end{figure}

Figure~\ref{fig2} shows the distributions of the velocity dispersion for the remaining targets (240 stars). Their 
velocity dispersions have been quadratically corrected for their mean photon noise for building the displayed distributions. 
The remaining velocity dispersion sources present in these distributions are: the instrumental error, the stellar jitter, 
the stellar oscillations, the non-detected (or not yet characterized) light planets and the non-detected blended spectra. 
Non-photonic error contributions have been randomly generated in our simulations using these two distributions.

\subsection{Impact of colour, metallicity and rotation}
\label{bvfeh}

\begin{figure}[!t]
  \begin{center}
    \epsfig{file=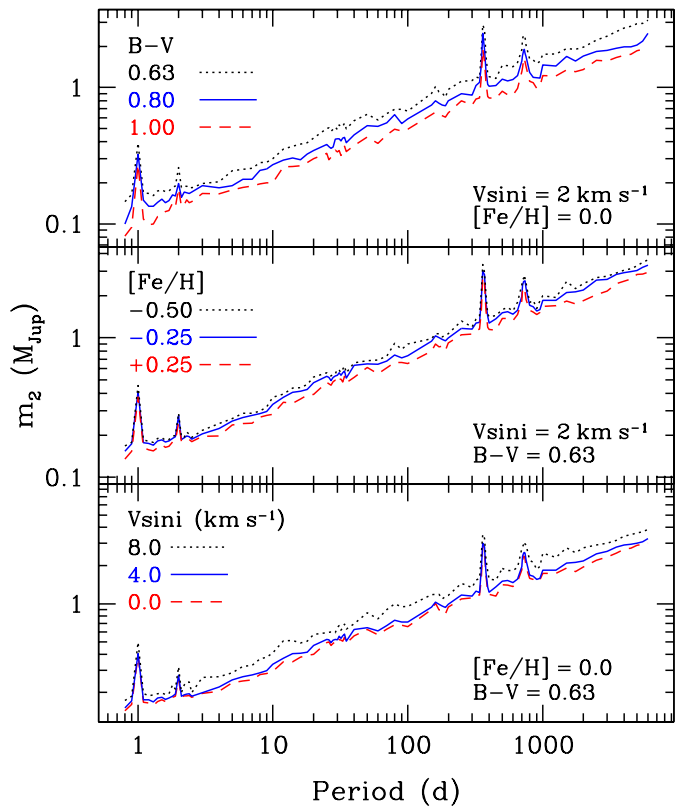, width=\hsize}
  \end{center}
  \caption{Impact of stellar properties on the 90\% detection limit of one sample star: $B-V$ (top), $[$Fe/H$]$ (middle) 
  and $v\sin i$ (bottom). An almost negligible impact of rotation and a weak impact of metallicity are observed. The moderate 
  impact of colour is mostly due to the differences in primary mass.\label{fig3}}
\end{figure}

Figure~\ref{fig3} shows the impact of the $B-V$ colour index, the metallicity $[$Fe/H$]$ and the projected rotational 
velocity $v\sin i$ on the 90\% detection limit obtained using the measurement dates and signal-to-noise ratios obtained for 
one of our sample stars. The photon noise errors have been computed using cross-correlation function parameters 
corresponding to the simulated stellar characteristics. 

The impact of rotation is negligible up to $v\sin i$ = 4\,km\,s$^{\rm -1}$, the value we have used for selecting our sample 
of slow rotators. The impact of metallicity is very weak whereas the impact of colour is higher but mostly due to primary 
mass differences. The absence of metallicity impact on the detection limits further demonstrates that the observed difference 
between the metallicity distributions for stars with and without planets (\cite{Santos01}, \cite{Santos03}, \cite{Santos04}) 
does not result from an observational bias.

\subsection{Results}
\label{results}

\begin{figure}[!t]
  \begin{center}
    \epsfig{file=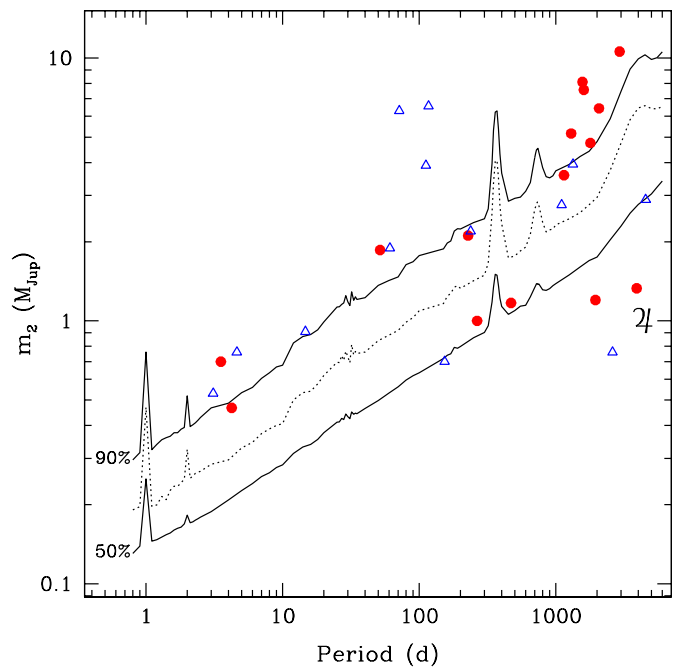, width=\hsize}
  \end{center}
  \caption{ELODIE 50 and 90\% global detection limits (solid lines). The dotted curve is the 90\% detection limit obtained 
  without including non-photonic error sources except the 6.5\,m\,s$^{\rm -1}$ instrumental error. This illustrates the 
  dramatic impact of these error sources (in particular the radial-velocity jitter induced by stellar activity). Planets 
  detected around our sample stars are noted by filled dots. The open triangles represent planet detected with ELODIE around 
  stars outside our sample or detections confirmed with ELODIE. The position of Jupiter is indicated with its symbol.
  \label{fig4}}
\end{figure}

Figure~\ref{fig4} shows the 50 and 90\% typical detection limits for the {\footnotesize ELODIE} survey. The dotted curve is 
the 90\% detection limit we would obtain without the presence of any non-photonic error source but the instrumental error 
(here set to 6.5\,m\,s$^{\rm -1}$). The filled dots are the planet detected around stars in our sample. Open triangles are 
used for planets detected with {\footnotesize ELODIE} outside our sample and for planets confirmed with 
{\footnotesize ELODIE}. The main trend of these curves is proportional to $P^{\rm 1/3}$ as expected from Kepler's laws. We 
note the enormous sensitivity decrease at $P$\,=1 and 2\,days and \,1 and 2\,years.

\begin{figure}[!t]
  \begin{center}
    \epsfig{file=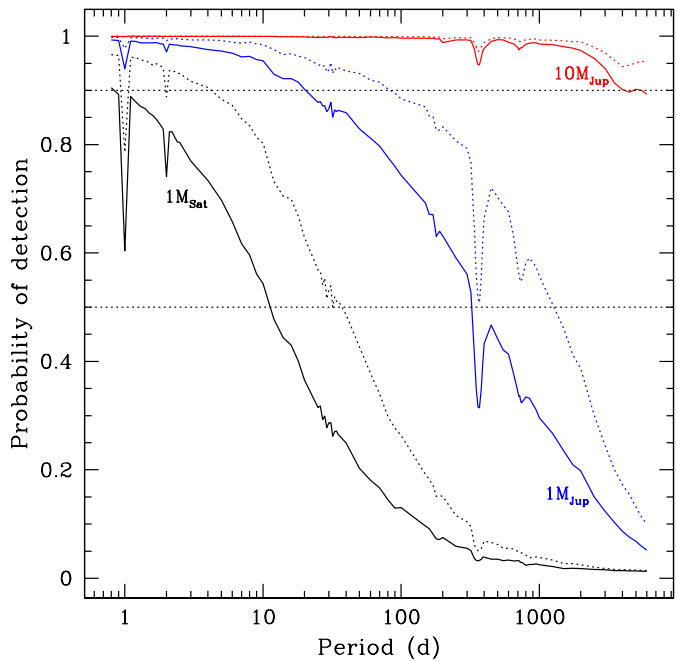, width=\hsize}
  \end{center}
  \caption{ELODIE detection probability versus orbital period for various companion masses (solid curves). The dotted curves 
  represent the detection probabilities obtained without accounting for non-photonic error sources.\label{fig5}}
\end{figure}

Probabilities of detection versus orbital period for different secondary masses (1\,M$_{\rm Sat}$, 1\,M$_{\rm Jup}$ and 
10\,M$_{\rm Jup}$) are displayed in Fig.~\ref{fig5}. The 
presence of non-photonic error sources is taken into account. 90\% of the Jupiter-mass planets are detected up to 
$P$\,$\simeq$\,20\,d and 50\% up to $P$\,= 300\,d. The probability of detection for brown-dwarf companions is above 90\% 
for all the period interval. The daily and yearly features are also clearly visible here. The dotted curves are obtained 
without including the contribution of non-photonic error sources except for the 6.5\,m\,s$^{\rm -1}$ instrumental error. 
These curves clearly show the dramatic impact of stellar jitter (and other error sources) on the detection sensitivity.
 
We find no sensitivity decrease between 1 and 2 days. Thus, planets in this period range, the so-called  "very hot Jupiters" 
(see e.g. \cite{Konacki03}, \cite{Bouchy04}, \cite{Konacki04}), would be easily detected if present around  our sample stars.

\section{Fraction of stars hosting a giant planet}

Following our obtained {\footnotesize ELODIE} detection limits, we can correct our effective detections for all the 
observational biases. We can also derive, by inverting the detection probability map, the fraction $f$ of stars in our sample 
hosting at least one giant planet (the outer planets of the two systems are not considered here) with 
$m_{\rm 2}$\,$\geq$\,0.47\,M$_{\rm Jup}$ and for different period intervals. We find:

\vspace{0.5cm}
\begin{center}
\mbox{}
\noindent\setlength{\fboxrule}{2pt}\setlength{\fboxsep}{3mm}\fbox{\parbox[t]{6cm}{
{\begin{center}\large\bf $f$\,=\,0.7\,$\pm$\,0.5\% for $P$\,$<$\,5\,d\end{center}}
{\begin{center}\large\bf $f$\,=\,4.0\,$\pm$\,1.1\% for $P$\,$<$\,1500\,d\end{center}}
{\begin{center}\large\bf $f$\,=\,7.3\,$\pm$\,1.5\% for $P$\,$<$\,3900\,d\end{center}}
}}
\end{center}
\vspace{0.5cm}

Details about our numerical simulations and these results on the global ELODIE survey sensitivity will be published in a 
forthcoming paper (Naef et al. 2004 in prep.).

\begin{acknowledgements}

We acknowledge support from the Swiss National Research Found ({\footnotesize FNS}), the Geneva University and the French 
{\footnotesize CNRS}. We are grateful to the Observatoire de Haute-Provence for the generous time allocation and for the 
constant support during the last ten years. 

\end{acknowledgements}

\end{document}